\documentclass{PoS}
\usepackage{lineno}
\usepackage[hyphens]{url}

\title{Constraining the Origin of Local Positrons with HAWC TeV Gamma-Ray Observations of Two Nearby Pulsar Wind Nebulae}

\ShortTitle{HAWC Nearby Pulsars}

\author{\speaker{F. Salesa Greus} $^a$, S. Casanova$^{a,c}$, B. Dingus$^b$, R. L\'{o}pez-Coto$^c$, and H. Zhou$^{b}$\\
\llap{$^a$}Instytut Fizyki J\k{a}drowej im. Henryka Niewodnicza\'{n}skiego, Polskiej Akademii Nauk\\
ul. Radzikowskiego 152, 31-342 Krakow, Poland\\
\llap{$^b$}Physics Division, Los Alamos National Laboratory\\
Los Alamos, NM 87545, USA\\
\llap{$^c$} Max-Planck-Institut f\"{u}r Kernphysik\\
Saupfercheckweg 1, Heidelberg, BW 69117, Germany\\

E-mail: \email{fsalesa@ifj.edu.pl}, \email{sabrina.casanova@ifj.edu.pl}, \email{dingus@lanl.gov}, \email{rlopez@mpi-hd.mpg.de}, \email{hzhou1@mtu.edu}}

\author{for the HAWC Collaboration\footnote{Complete list of authors at http://www.hawc-observatory.org/collaboration/icrc2017.php}}

\abstract{The HAWC Gamma-Ray Observatory has reported the discovery of TeV gamma-ray emission extending several degrees around the positions of Geminga and B0656+14 pulsars. Assuming these gamma rays are produced by inverse Compton scattering off low-energy photons in electron halos around the pulsars, we determine the diffusion of electrons and positrons in the local interstellar medium. We will present the morphological and spectral studies of these two VHE gamma-ray sources and the derived positron spectrum at Earth.}

\FullConference{35th International Cosmic Ray Conference - ICRC217 -\\
		10-20 July, 2017\\
		Bexco, Busan, Korea}

\begin{document}

\section{Introduction}

An unexpected excess in the local positron flux of energies above 10 GeV was detected by the PAMELA satellite in 2009~\cite{PAMELA}. This excess was confirmed by Fermi-LAT~\cite{Fermi}, and more recently the AMS-02 satellite provided high precision measurements and also extended them to higher energies~\cite{AMS}.
Presently, there is no clear explanation why the positron flux is higher than what would be expected from conventional models of cosmic ray propagation~\cite{Moska}. Some authors claim that it can be the result of nearby dark matter interactions~\cite{Ibarra}. Another hypothesis is that nearby cosmic-ray accelerators, like pulsar wind nebulae (PWNe), could be actually the sources of these extra positrons~\cite{Blasi}.

Among the PWNe candidates the Geminga pulsar has been suggested as one of the best candidates since it is close to the Earth and old enough to be able to produce high energy electrons and positrons that would contribute significantly to the excess observed by the satellites. The observations of an extended TeV gamma-ray emission around the Geminga pulsar by Milagro~\cite{Milagro} were used by some authors to give extra support to the nearby source hypothesis~\cite{Aharonian1,Yuksel}.

Here we will report the latest results on the observations of Geminga by HAWC, together with the discovery of a similar extended emission from PSR B0656+14. The improvement on the HAWC performance with respect to Milagro provides a more accurate measurement of the gamma-ray flux and morphology from both sources, allowing us to constrain the origin of local positrons.

\section{HAWC Observation}

The HAWC Gamma-Ray Observatory has been fully operational since March 2015~\cite{Crab}. In the previous ICRC the observation of the Geminga extended TeV emission was presented~\cite{Geminga2015} using a dataset which mostly included data from a partial configuration of the HAWC array. This observation constituted the first confirmation of the Geminga detection reported by Milagro experiment.  Currently, with 17 months of data with the complete HAWC array a new source catalog was released~\cite{Catalog}. In the catalog, the detection of extended emission in the vicinity of Geminga and PSR B0656+14 was reported associated with the HAWC source 2HWC J0635+180 and 2HWC J0700+143.

\begin{figure}[htpb]
\begin{center}
\begin{tabular}{c c}
\hspace{-0.2 cm}
\includegraphics[width=0.5\linewidth, angle=0]{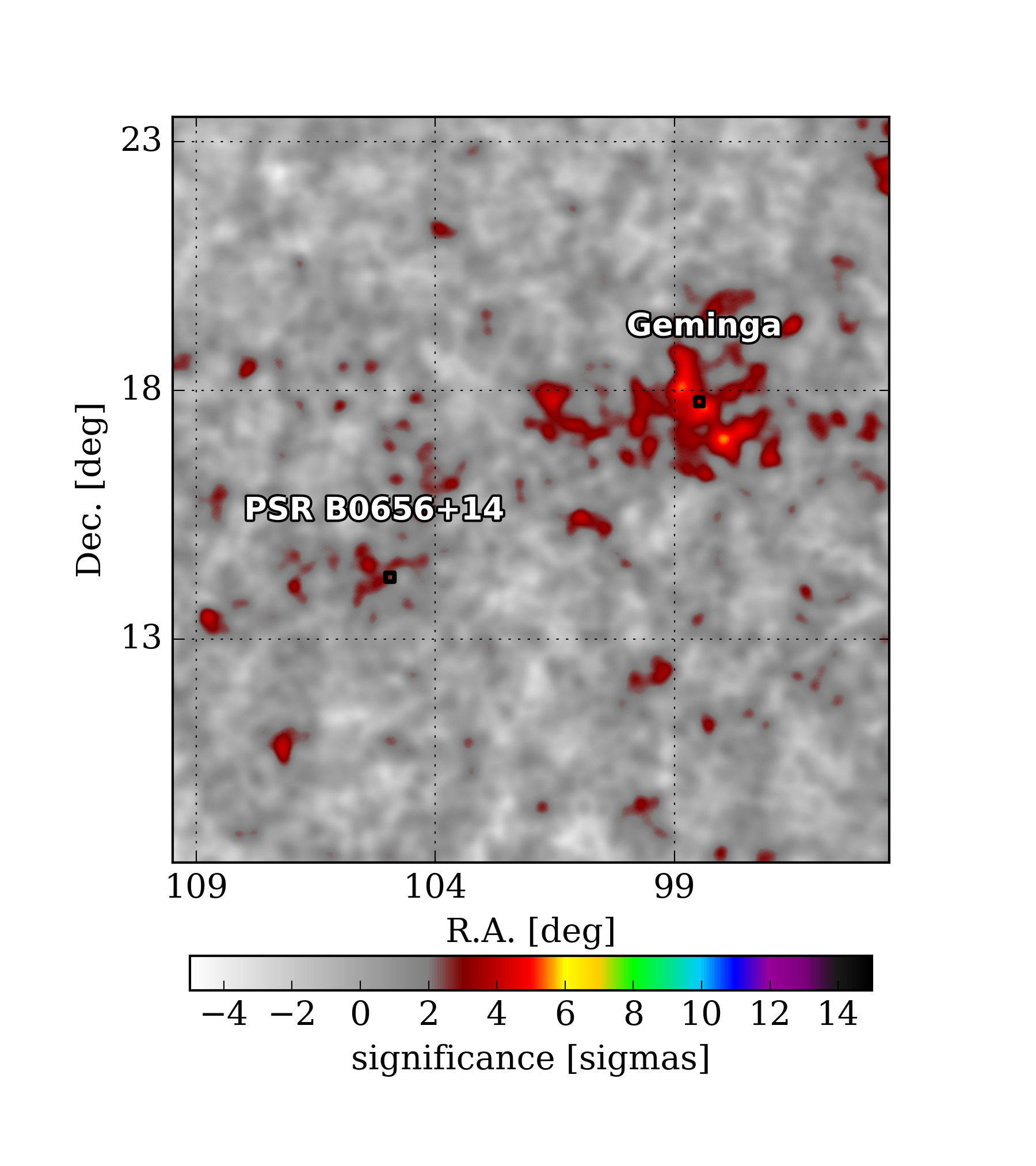}
\hspace{0.0 cm}
\includegraphics[width=0.5\linewidth, angle=0]{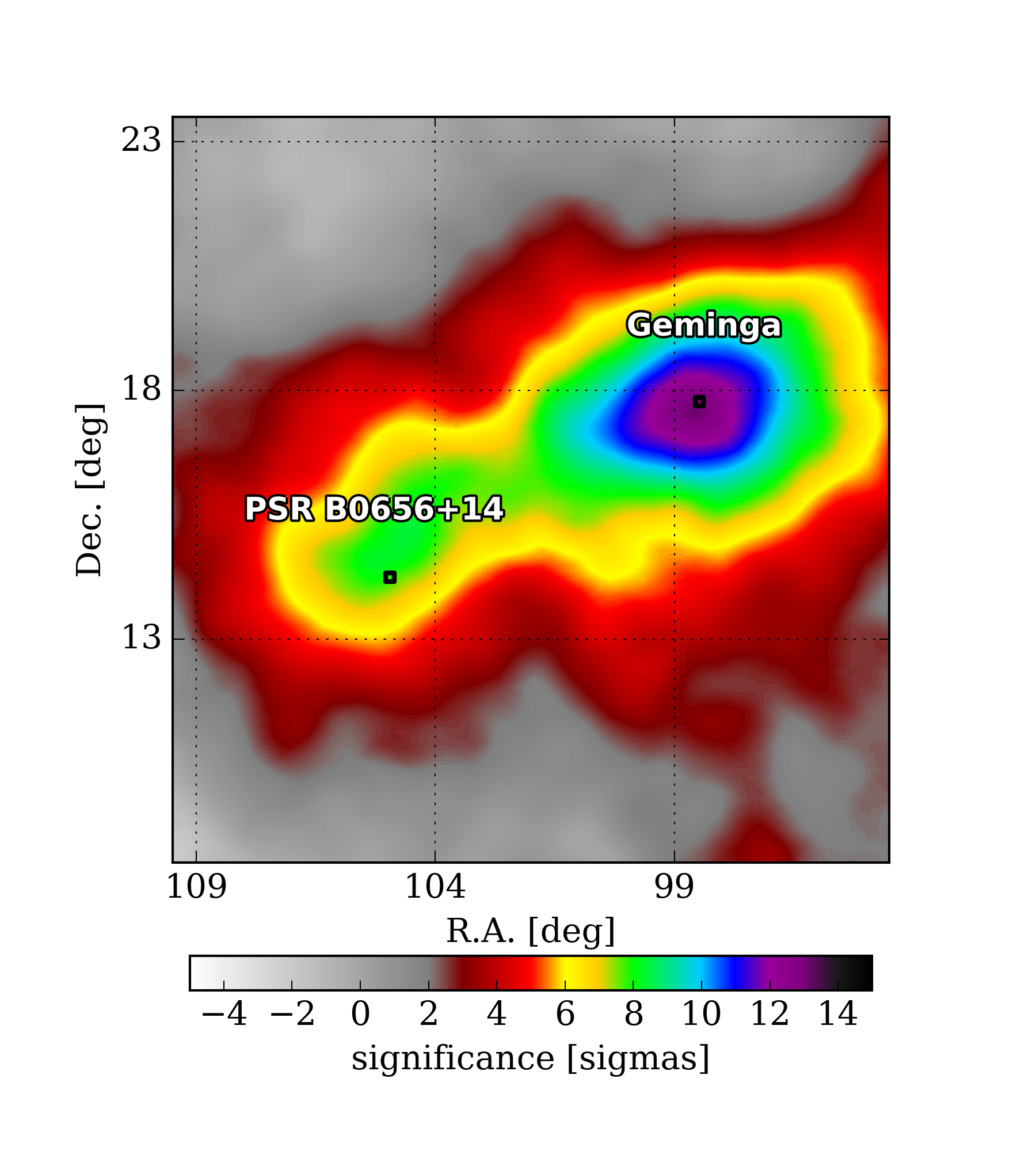}
\end{tabular}
\caption{Significance map for the Geminga and PSR B0656+14 region convolved with the point spread function (left) and with the morphology which assumes particle diffusion (right).}

\label{fig: map}
\end{center}
\end{figure}

A morphological analysis was performed on these two sources (described briefly in the next section). The results show that the emission from the Geminga location reaches 13.1 standard deviations when a diffusion model morphology is used (see Fig.~\ref{fig: map}). In addition, a similar extended TeV gamma-ray emission in the vicinity of PSR B0656+14 has been detected at 8.1 sigma level. Unlike Geminga, the extended emission of PSR B0656+14 was not reported before by any gamma-ray experiment.

The Geminga pulsar (PSR J0633+1746) is an isolated neutron star in the Gemini constellation. The estimated age of the pulsar is 3.42$\times 10^{5}$ yr and the distance of 250 pc with a spin-down luminosity of 3.26$\times 10^{34}$ erg s$^{-1}$~\cite{ATNF}.

The pulsar B0656+14 is also an isolated pulsar which has been associated with the large (25 degree) supernova remnant covering the Gemini and Monoceros constellations called the ``Monogem ring''. Some authors speculated that the Monogem Ring SNR could be the long-searched-for source of the highest energy protons producing the ``knee'' feature in the cosmic-ray spectrum~\cite{Monogem}.

The characteristics of PSR B0656+14 are similar to those of Geminga, i.e, the estimated age is 1.1$\times 10^{5}$ yr, the distance to Earth is 288 pc, and the spin-down luminosity is 3.8$\times 10^{34}$ erg s$^{-1}$~\cite{ATNF}.

\section{Morphological studies}

\begin{figure}[htpb]
\begin{center}
\includegraphics[width=0.75\linewidth, angle=0]{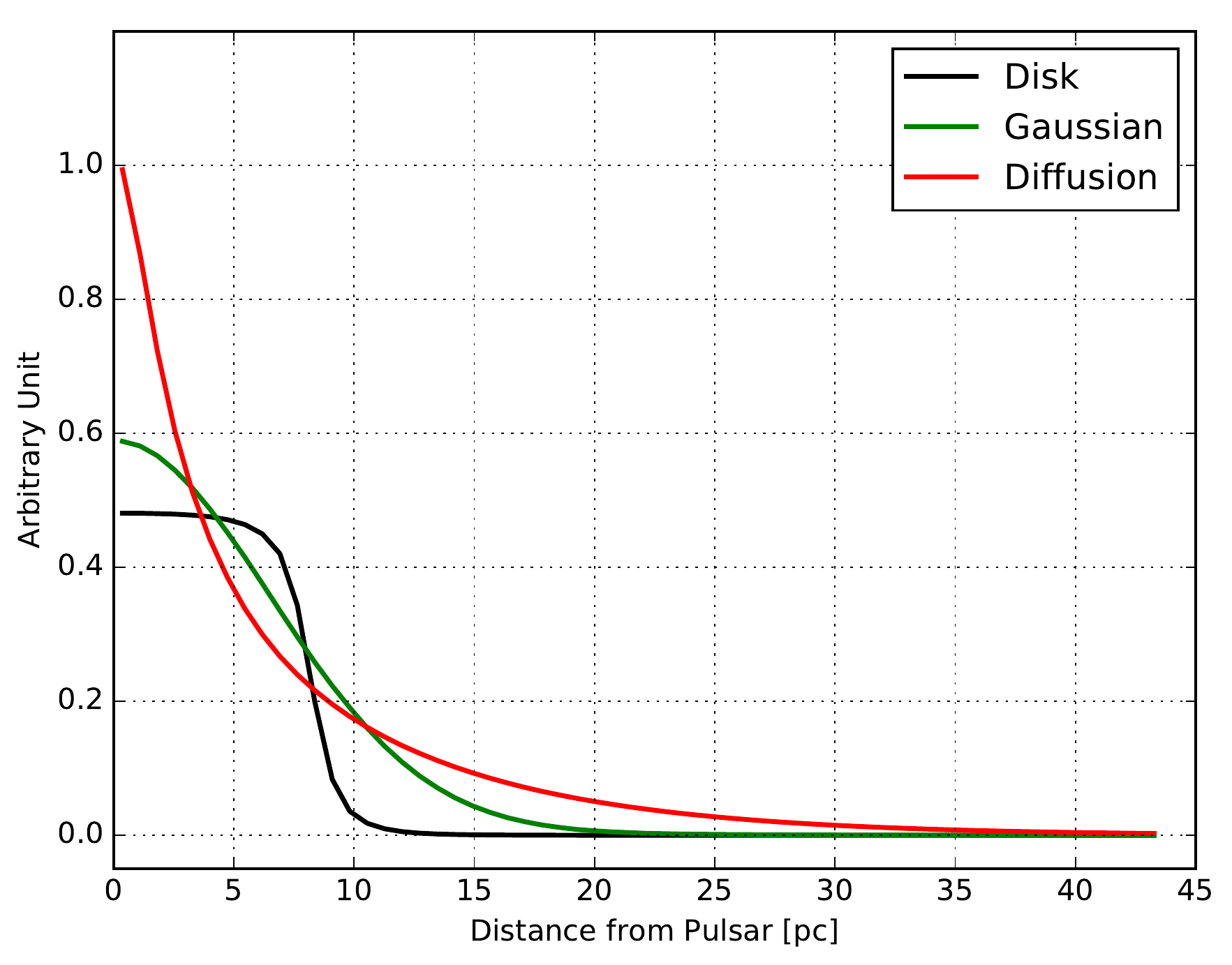}
\caption{Geminga flux per surface unit for different radial distances. Overlaid are the three morphological models tested.}
\label{fig: morph}
\end{center}
\end{figure}

For the analysis, three morphology assumptions were tested (see Fig.~\ref{fig: morph}). The first morphology approximated the source as extended disks where a constant (average) flux of gamma-rays fitted within a distance from the source center location. The second morphology is a symmetrical two-dimensional Gaussian. In this case, the fitted flux of gamma-rays depends on the distance from the source center location as following a 2D-Gaussian. Apart from these two morphologies, which do not depend on physical assumptions, a third one based on~\cite{Aharonian} assumes electron and positron pairs diffusing into the interstellar medium around the pulsar, producing gamma rays through inverse Compton scattering of CMB photons.

More details on the morphology studies can be found elsewhere in the proceedings for this conference~\cite{Hao}.

\section{Spectral studies}

A first estimate of the source gamma-ray flux was provided in the recent HAWC catalog~\cite{Catalog}. The most updated result using a diffusion model morphology will be presented at the conference.

\section{Discussion}

HAWC observations confirmed the extended TeV emission from the Geminga region. In addition, a similar extended emission was reported from the vicinity of PSR B0656+14. Both are promising candidates to be the sources of the positron excess above 10 GeV detected at the Earth and reported by several satellites. Both morphological and spectral studies have been carried out using the 17 moths of HAWC dataset.

These studies provide information about the diffusion coefficient and gamma-ray spectrum which can be used as input parameters for the EDGE code~\cite{Ruben}. EDGE takes into account the theory of electron diffusion, and computes the expected positron flux produced at the Earth by a pulsar.

The derived positron spectrum at Earth considering the measurements of Geminga and PSR B0656+14 by HAWC will be presented at the conference.

\section*{Acknowledgments}
\footnotesize{
We acknowledge the support from: the US National Science Foundation (NSF); the
US Department of Energy Office of High-Energy Physics; the Laboratory Directed
Research and Development (LDRD) program of Los Alamos National Laboratory;
Consejo Nacional de Ciencia y Tecnolog\'{\i}a (CONACyT), M{\'e}xico (grants
271051, 232656, 260378, 179588, 239762, 254964, 271737, 258865, 243290,
132197), Laboratorio Nacional HAWC de rayos gamma; L'OREAL Fellowship for
Women in Science 2014; Red HAWC, M{\'e}xico; DGAPA-UNAM (grants IG100317,
IN111315, IN111716-3, IA102715, 109916, IA102917); VIEP-BUAP; PIFI 2012, 2013,
PROFOCIE 2014, 2015;the University of Wisconsin Alumni Research Foundation;
the Institute of Geophysics, Planetary Physics, and Signatures at Los Alamos
National Laboratory; Polish Science Centre grant DEC-2014/13/B/ST9/945;
Coordinaci{\'o}n de la Investigaci{\'o}n Cient\'{\i}fica de la Universidad
Michoacana. Thanks to Luciano D\'{\i}az and Eduardo Murrieta for technical support.}

\end{document}